\documentclass[sigconf,screen]{acmart}
\AtBeginDocument{%
  \providecommand\BibTeX{{%
    \normalfont B\kern-0.5em{\scshape i\kern-0.25em b}\kern-0.8em\TeX}}}

\usepackage{listings}

\lstdefinestyle{yaml}{
     basicstyle=\color{blue}\footnotesize,
     rulecolor=\color{black},
     string=[s]{'}{'},
     stringstyle=\color{blue},
     comment=[l]{:},
     commentstyle=\color{black},
     morecomment=[l]{-},
     frame=single,
     numberstyle=\tiny\color{gray},
     numbersep=5pt,
     captionpos=b,
     numbers=left,
     breakatwhitespace=false,
     float,
     floatplacement=tbp,
     linewidth=0.99\linewidth
}

\setcopyright{rightsretained}
\acmPrice{}
\acmDOI{10.1145/3617574.3617859}
\acmYear{2023}
\copyrightyear{2023}
\acmSubmissionID{fsews23sesafemlmain-id5149-p}
\acmISBN{979-8-4007-0379-9/23/12}
\acmConference[SE4SafeML '23]{Proceedings of the 1st International Workshop on Dependability and Trustworthiness of Safety-Critical Systems with Machine Learned Components}{December 4, 2023}{San Francisco, CA, USA}
\acmBooktitle{Proceedings of the 1st International Workshop on Dependability and Trustworthiness of Safety-Critical Systems with Machine Learned Components (SE4SafeML '23), December 4, 2023, San Francisco, CA, USA}
\received{2023-07-04}
\received[accepted]{2023-08-10}

\begin{document}

\title{MLGuard: Defend Your Machine Learning Model!}

\author{Sheng Wong}
\email{wongsh@deakin.edu.au}
\author{Scott Barnett}
\email{scott.barnett@deakin.edu.au}

\affiliation{%
  \institution{Deakin University}
  \city{Melbourne}
  \state{Victoria}
  \country{Australia}
  \postcode{3125}
}

\author{Jessica Rivera-Villicana}
\email{jessica.rivera.villicana@rmit.edu.au}
\affiliation{%
  \institution{RMIT University}
  \city{Melbourne}
  \state{Victoria}
  \country{Australia}
  \postcode{3000}
}

\author{Anj Simmons}
\author{Hala Abdelkader}
\email{a.simmons@deakin.edu.au}
\email{h.abdelkader@deakin.edu.au}
\affiliation{%
  \institution{Deakin University}
  \city{Melbourne}
  \state{Victoria}
  \country{Australia}
  \postcode{3125}
}

\author{Jean-Guy Schneider}
\email{Jean-Guy.Schneider@monash.edu}
\affiliation{%
  \institution{Monash University}
  \streetaddress{221 Burwood Highway}
  \city{Clayton}
  \state{Victoria}
  \country{Australia}
  \postcode{3800}
}

\author{Rajesh Vasa}
\email{rajesh.vasa@deakin.edu.au}
\affiliation{%
  \institution{Deakin University}
  \city{Melbourne}
  \state{Victoria}
  \country{Australia}
  \postcode{3125}
}

\renewcommand{\shortauthors}{Wong et al.}

\begin{abstract}
Machine Learning (ML) is used in critical highly regulated and high-stakes fields such as finance, medicine, and transportation. The correctness of these ML applications is important for human safety and economic benefit. Progress has been made on improving ML testing and monitoring of ML. However, these approaches do not provide i) pre/post conditions to handle uncertainty, ii) defining corrective actions based on probabilistic outcomes, or iii) continual verification during system operation.  
In this paper, we propose MLGuard, a new approach to specify contracts for ML applications. Our approach consists of a) an ML contract specification defining pre/post conditions, invariants, and altering behaviours, b) generated validation models to determine the probability of contract violation, and c) an ML wrapper generator to enforce the contract and respond to violations. Our work is intended to provide the overarching framework required for building ML applications and monitoring their safety.
\end{abstract}




\maketitle

\section{Introduction}


Robustness in business software where the domain is well understood is achieved  through software testing, and adherence to best practices and processes. However, for Machine Learning (ML) systems this is insufficient. ML systems are dependent on data input streams that are non-stationary. As a result, ML is behaviour is underspecified \cite{damour2022underspecification} in the presence of subtle changes in the data (i.e. data shift \cite{moreno2012}). Data schema validation alone is insufficient as detecting violations of these conditions, e.g., out of distribution data \cite{hendrycks2017}, can only be done probabilistically. We hypothesise that robustness can be incrementally realised in the context of ML through an interface specification (contract) that a) operates on point-estimates and distributions, b) encapsulates modelling assumptions, and c) models uncertainty as a first class citizen.




To achieve robustness for ML, research has focused on testing against noisy or malicious input data \cite{applis2021assessing, gao2020fuzz,zhang2020uncertainty, kwiatkowska2020safety}. However, the world is non-stationary and assumptions made offline are not guaranteed to hold during system operation. In addition, runtime validation is required to handle ML specific failure modes (i.e. miscalibration~\cite{guo2017calibration} and performance degradation under data shift~\cite{moreno2012}) independent to how the model responds to adversarial examples. Our research provides the constructs and plumbing for software engineers to leverage existing algorithms that detect these failure modes (due to the difficulty of detecting these failure modes, algorithms in turn may require additional ML models to perform the validation).

\begin{figure*}[t]
    \centering
    \includegraphics[width=0.57\linewidth]{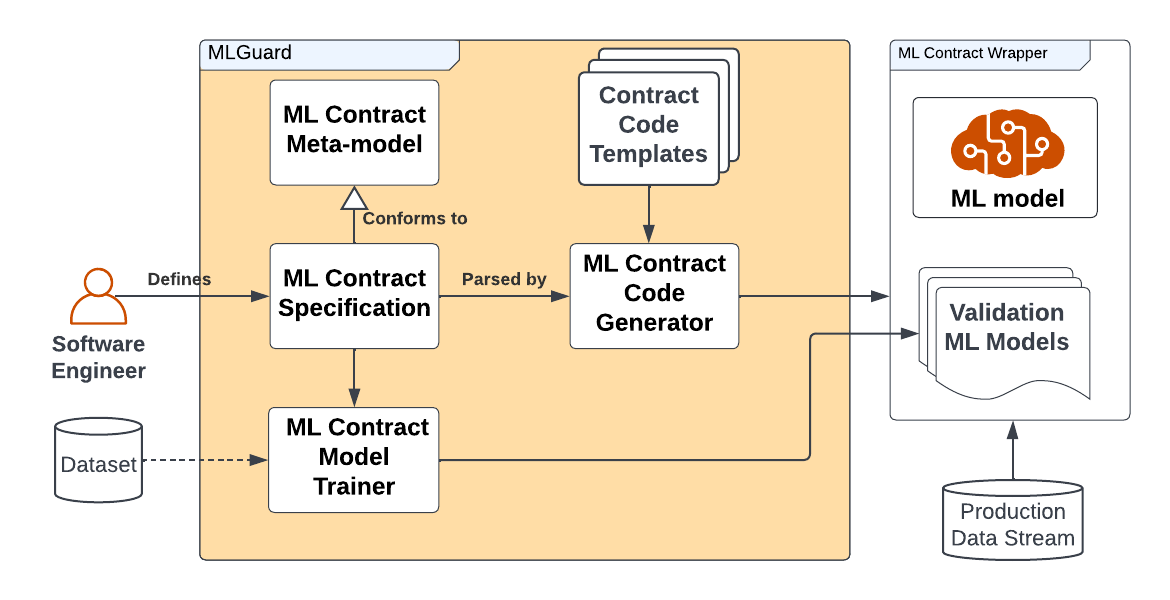}
    \caption{Our approach consisting of 1) an ML contract specification, 2) an ML contract model trainer to generate validation ML models that determine the probability of contract violations, and 3) a generated wrapper to defend models from contract violations and trigger contract violation handling logic if violations occur.}
    \label{fig:approach}
\end{figure*}


Studies have managed to produce data validation tools for ML systems that continuously monitor data that are fed into the system. Data Linter~\cite{Hynes2017a} acts as simplified and general validation tool that can be used to automatically inspect data and detect miscoding lints, outliers and scaling issues~\cite{Hynes2017a}. Data validation tools developed in-house like Deequ~\cite{Schelter2018, schelter2019} and Google's TFX allow user-specified constraints~\cite{Schelter2018, Baylor2017}. Most of the validation tools available use constraints and conditions that are: deterministic in nature, provide partial support for ML failure modes, and typically act as a warning system for data with anomalous or invalid characteristics. In addition, data validation tools presented are insufficient in producing a reliable and robust ML system, since the ``correctness'' of data is variable depending on the context and scenario of the ML system.



Best practice recommends setting up alerting and monitoring infrastructure \cite{Breck2017, Schelter2018, Baylor2017} and ML specific architectural tactics \cite{Cummaudo2020b}. However, how to specify the conditions to monitor and the actions to take when a condition fails is left to the developer. ML toolsuites for validation (e.g., Tensorflow Data Validation) provide a set of validators and abstractions for their specification. However, these tools only offer partial support for ML failure modes and automation. Specifically, Tensorflow supports data schema and constraint suggestions, but leave more sophisticated validation checks such as out of distribution detection to the developer to implement and configure.




Inspired by concepts from design by contract~\cite{meyer1992applying}, we propose a new approach, MLGuard, for specifying and validating ML contracts.
Our approach takes an ML Contract Specification and generates an ML wrapper with \textit{both the code and trained ML models for validating ML contracts}. We propose an ML contract specification language with i) ML specific concepts (e.g., `uncertain') and ii) actions to take when the ML contract fails (e.g., log warning, throw exception, propagate uncertainty).
To the best of our knowledge, this is the first proposed approach for specifying and validating probabilistic contracts for ML. Although no formal guarantees can be made of the absolute safety of the ML system, our approach provides a structured semi-automated way to help developers work towards improving the safety of the ML applications they develop by automatically detecting and responding to contract violations. MLGuard is designed to provide the scaffolding for specifying and validating contracts that, optionally, include additional validation algorithms i.e. a data drift detector.

\section{Motivation}

To motivate our work, consider the case of software for automated epileptic seizure detection---classifying a segment of electroencephalogram (EEG) data as seizure or non-seizure. While a large number of papers have developed ML models for this task \cite{shoeibi2021epileptic}, these models come with (often undocumented) conditions that must be satisfied for the output produced by the model to be reliable. This poses a safety concern as over-interpretation of EEG data (false positives) might lead to incorrect diagnosis and treatment, and this causes numerous medication side-effects, driving restrictions, increased chance of mental illness and discrimination of job opportunities while under-diagnosis (false negatives) causes delayed treatment which might increase the risk of mortality and other physical injuries.

We elaborate on these challenges below, and propose a vision in section~\ref{section:vision} to address them, thereby improving the safety of ML models when deployed in real-world applications.


\textbf{No machine-checkable ML specification language.} Well-designed software components document inputs, outputs, types, pre/post conditions to be satisfied, and exceptions that may be raised. However, ML models lack a full machine-checkable specification. For example, even though an ML model accepts a vector with the same type and dimensions as the EEG data, this does not necessarily mean that the model is a suitable choice. To determine if the model is compatible with the application, one also needs to consider whether the statistical characteristics of the training data and modelling assumptions match those of the application domain.

To assist in assessing the suitability of an ML model, proposals have been made for standardised documentation templates (i.e. data sheets~\cite{gebru2021datasheets} and model cards~\cite{mitchell2019}). However, documentation templates require users of the ML model to manually read and interpret the ML model documentation (if any) without providing any machine-verifiable rules for safe use.


\textbf{No mechanism to express uncertainty in validation rules.} Different electrode placements, sampling frequencies, and filtering are possible. If these do not match those of the data the ML model was trained on, the ML model will still produce a result, but it cannot be trusted. The patient demographics may also affect the accuracy of the model. For example, it should not be assumed that a model trained on EEG from adults will work as well on EEG from children.

To ensure that the EEG data is compatible with the data on which the model was trained, one can make use of an out of distribution detection model to determine whether the assumptions of the model have been violated, i.e. the serving data in production should be distributed similarly to the data the model was trained on. However, this violation can only be detected probabilistically rather than with absolute certainty.

The conditions for these contracts operate in high-dimensional latent spaces. For ML, the input and learned latent spaces are where pre/post conditions are required to verify system behaviour. Currently we lack the mathematical constructs for guaranteeing behaviour across a high-dimensional latent space. 


\textbf{It is unclear how the application should respond to probabilistic violations}. Unlike traditional software, ML behaviour is dependent on training data---demonstrating correctness offline is no guarantee of the system's operating behaviour online. Thus violations of contracts need to be detected and responded to during operation rather than at design time. Best practice recommends setting up existing alerting and monitoring infrastructure \cite{Breck2017}. However, what is not specified is 1) how to configure alarms and alerts, and 2) how the system should respond.

\section{A Vision for ML Contracts}
\label{section:vision}
We propose MLGuard for automatically validating whether incoming data conforms to an ML contract and handling violations. MLGuard is a practical approach for dealing with the limitations outlined in the Motivation. An overview of our proposed approach is presented in Fig.~\ref{fig:approach}, and more detailed descriptions of each MLGuard component are provided below.


\subsection{ML Contract Meta-model and Specification}
The ML Contract Meta-model provides the abstractions needed to specify ML contracts, serving as the basis for a \textbf{machine-checkable ML specification language}. For example, in addition to validating the data schema, one can specify that the model requires input data to be distributed similarly to the training data. The meta-model also provides the abstractions to define strategies for detecting probabilistic violations and for responding to probabilistic violations (the components to support this are elaborated on in the following sections).

We borrow the concept of declarative definition of constraints developed by Deequ~\cite{Schelter2018, schelter2019}, and extend these ideas to allow for probabilistic conditions and specify actions to take. This approach will enable specifications be made regarding i) probabilistic conditions on inputs, ii) what methods will be used to detect probabilistic violations, and iii) approaches for dealing with probabilistic outcomes. For example, what should happen when conditions are violated with a confidence of 55\% produced by a ML system. Software engineers write an ML Contract Specification tailored to their application needs that instantiates concepts in the meta-model. A sample contract is provided in Listing \ref{lst:demostration} using a YAML based syntax, but in future we will explore use of domain specific languages and fluent APIs.

\begin{lstlisting}[style=yaml,caption={Sample ML Contract Specification for seizure detection ML model. The ML contract allows specifying preconditions making use of probabilistic concepts (e.g., distributions match), methods for detecting probabilistic violations (e.g., out of distribution detection), and how to respond to probabilistic violations (e.g., whether to log a warning or raise an exception).},label={lst:demostration}]
Contract:
   Model:
      Name: seizure_detection_ml_model
      Location: /pretrained/seizure_model.onnx
      Documentation: /doc/seizure_model_card.html
   Data:
      - input_steam
      - output_stream
      - /data/eeg_train
   Preconditions:
      Distribution_Matches:
         DatasetA: input_steam
         DatasetB: /data/eeg_train
         Validation_model:
            Type: out_of_distribution_detector
            Method: likelihood_ratios_for_ood
         Trigger_conditions:
            Confidence_threshold: 0.95
         Action_if_violated: log_warning
      Schema_Matches:
         Dataset: input_steam
         Schema: /schema/eeg-10-20-system-256hz
         Action_if_violated: exception
   Postconditions:
      Probabilities_sum_to_one:
         Dataset: output_stream
         Action_if_violated: exception
\end{lstlisting}

\subsection{ML Contract Model Trainer}
Validating compliance with the ML Contract Specification requires Validation ML Models to detect probabilistic violations of contracts. The type of Validation ML Model to use may be specified as part of the ML Contract Specification, along with configurable thresholds at which to trigger contract violation handling logic, which together form a \textbf{mechanism to express uncertainty in validation rules}. For example, to validate that an instance of the input data is from the same distribution as the data the ML model was trained on, one may make use of an out of distribution detector. In the case of the Likelihood Ratios for Out-of-Distribution Detection method \cite{ren2019likelihood}, this requires training deep-generative models to determine the probability that the data is out of distribution and correct for background statistics.

The role of the ML Contract Model Trainer is to automatically train the Validation ML Models (not to be confused with the ML Model that they are guarding) according to the configuration provided in the ML Contract Specification. To support software engineers uncertain about which type of ML Validation Model to use and how to configure it, we intend to explore approaches based on AutoML~\cite{he2021automl} to automatically select and train appropriate Validation ML Models to enforce the constraints in the ML Contract Specification when the Validation ML Model to use is left unspecified. Our approach will also allow for threshold conditions at which to trigger a violation to be automatically learned from data and refined based on user feedback in the case the the software engineer is uncertain about which threshold value to specify.

\subsection{ML Contract Wrapper}
A standard approach to addressing the issue of robustness is to introduce a wrapper \cite{shahrokni2013systematic} to guard against invalid behaviour. The ML Contract Code Generator selects and instantiates Contract Code Templates with information in the ML Contract Specification. The generated wrapper code includes the i) trained ML model, ii) Validation ML Models for use in pre/post conditions, and iii) code for checking pre/post conditions (using the Validation ML Models) to guard the trained model and trigger contract violation handling logic to \textbf{respond to probabilistic violations}.
The wrapper can be configured (via the ML Contract Specification) to respond to contract violations in a manner appropriate to the application and nature of violation. For example, should an exception be thrown, error messages logged, or uncertainty be propagated through the system?

\section{Research Questions}
The work proposed and discussed in the previous sections led us to pose the following research questions. Our plan to answering them is further discussed in section~\ref{section:futureplans}.

\begin{itemize}
    \item \textbf{RQ1} What are the abstractions required for specifying ML contracts?
   \item \textbf{RQ2} What software architecture is required to enable the generation of a wrapper for enforcing ML contracts?
   \item \textbf{RQ3} How effective is an ML contract in practice?
\end{itemize}

\section{Future Plans}
\label{section:futureplans}

Our research will progress in three phases. Phase~1 will focus on extracting concepts for the ML contract meta-model from the literature and defining the ML contract specification language. Phase~2 will involve experimental evaluation of the the ML Contract. Finally, Phase~3 will investigate the effectiveness of MLGuard in an industry context.  

\textit{Phase 1: ML Contract meta-model and specification language:} To answer research question \textit{RQ1: What are the abstractions required for specifying ML contracts?} we will expand an existing specification language. We plan to follow an iterative approach inspired by Grounded theory \cite{stol2016grounded} to mine concepts from the literature (both academic and grey literature). The goal of this phase is to identify the concepts for specifying ML Contracts and defining a validation plan. The expected outcome from this phase will be 1) the ML Contract meta-model, 2) an ML Contract specification language, and 3) a set of ML specific conditions for validation.

\textit{Phase 2: Experimental evaluation of ML Contracts:} The next phase of research will involve developing a prototype of our solution to answer \textit{RQ2: What software architecture is required to enable the generation of a wrapper for enforcing ML contracts?} Concepts borrowed from Model Driven Engineering (MDE) will be applied to design the generators (code and trained models). Our experiment will evaluate ML Contracts against other automated and manual approaches to specifying validation logic for ML. The expected outcomes from this phase will be 1) a prototype tool MLGuard, 2) code templates for ML Contract wrappers, 3) configurations for training Validation ML models, and 4) a set of ML Wrappers generated for existing models. 
    
\textit{Phase 3: Industry case study:} To address \textit{RQ3: How effective is an ML contract in practice?} the final phase of the study will evaluate how our approach can be integrated into existing software engineering workflows. We plan to run a series of industry case studies where practitioners evaluate MLGuard on ML projects. The focus of the case studies will be to identify i) user-acceptance of the approach, ii) barriers to adoption, and iii) ongoing maintenance implications. 

\begin{acks}
The research was supported by a Deakin University Postgraduate Research Scholarship (DUPR) and a National Intelligence Postdoctoral Grant (NIPG-2021-006).
\end{acks}

\bibliographystyle{ACM-Reference-Format}
\bibliography{biblio.bib}


\begin{thebibliography}{22}


\ifx \showCODEN    \undefined \def \showCODEN     #1{\unskip}     \fi
\ifx \showDOI      \undefined \def \showDOI       #1{#1}\fi
\ifx \showISBNx    \undefined \def \showISBNx     #1{\unskip}     \fi
\ifx \showISBNxiii \undefined \def \showISBNxiii  #1{\unskip}     \fi
\ifx \showISSN     \undefined \def \showISSN      #1{\unskip}     \fi
\ifx \showLCCN     \undefined \def \showLCCN      #1{\unskip}     \fi
\ifx \shownote     \undefined \def \shownote      #1{#1}          \fi
\ifx \showarticletitle \undefined \def \showarticletitle #1{#1}   \fi
\ifx \showURL      \undefined \def \showURL       {\relax}        \fi
\providecommand\bibfield[2]{#2}
\providecommand\bibinfo[2]{#2}
\providecommand\natexlab[1]{#1}
\providecommand\showeprint[2][]{arXiv:#2}

\bibitem[Applis et~al\mbox{.}(2021)]%
        {applis2021assessing}
\bibfield{author}{\bibinfo{person}{Leonhard Applis}, \bibinfo{person}{Annibale
  Panichella}, {and} \bibinfo{person}{Arie van Deursen}.}
  \bibinfo{year}{2021}\natexlab{}.
\newblock \showarticletitle{Assessing Robustness of ML-Based Program Analysis
  Tools using Metamorphic Program Transformations}. In
  \bibinfo{booktitle}{\emph{2021 36th IEEE/ACM International Conference on
  Automated Software Engineering (ASE)}}. IEEE, \bibinfo{pages}{1377--1381}.
\newblock


\bibitem[Baylor et~al\mbox{.}(2017)]%
        {Baylor2017}
\bibfield{author}{\bibinfo{person}{Denis Baylor} {et~al\mbox{.}}}
  \bibinfo{year}{2017}\natexlab{}.
\newblock \showarticletitle{{TFX: A TensorFlow-Based Production-Scale Machine
  Learning Platform}}. In \bibinfo{booktitle}{\emph{Proceedings of the 23rd ACM
  SIGKDD International Conference on Knowledge Discovery and Data Mining}}.
  \bibinfo{publisher}{ACM}, \bibinfo{pages}{1387--1395}.
\newblock


\bibitem[Breck et~al\mbox{.}(2017)]%
        {Breck2017}
\bibfield{author}{\bibinfo{person}{Eric Breck}, \bibinfo{person}{Shanqing Cai},
  \bibinfo{person}{Eric Nielsen}, \bibinfo{person}{Michael Salib}, {and}
  \bibinfo{person}{D. Sculley}.} \bibinfo{year}{2017}\natexlab{}.
\newblock \showarticletitle{The {ML} Test Score: A Rubric for {ML} Production
  Readiness and Technical Debt Reduction}. In
  \bibinfo{booktitle}{\emph{Proceedings of IEEE Big Data}}.
\newblock


\bibitem[Cummaudo et~al\mbox{.}(2020)]%
        {Cummaudo2020b}
\bibfield{author}{\bibinfo{person}{Alex Cummaudo}, \bibinfo{person}{Scott
  Barnett}, \bibinfo{person}{Rajesh Vasa}, \bibinfo{person}{John Grundy}, {and}
  \bibinfo{person}{Mohamed Abdelrazek}.} \bibinfo{year}{2020}\natexlab{}.
\newblock \showarticletitle{{Beware the Evolving `Intelligent' Web Service! An
  Integration Architecture Tactic to Guard AI-First Components}}. In
  \bibinfo{booktitle}{\emph{Proceedings of the 28th ACM Joint Meeting on
  European Software Engineering Conference and Symposium on the Foundations of
  Software Engineering}}. \bibinfo{publisher}{ACM}, \bibinfo{pages}{269--280}.
\newblock


\bibitem[D'Amour et~al\mbox{.}(2022)]%
        {damour2022underspecification}
\bibfield{author}{\bibinfo{person}{Alexander D'Amour} {et~al\mbox{.}}}
  \bibinfo{year}{2022}\natexlab{}.
\newblock \showarticletitle{Underspecification Presents Challenges for
  Credibility in Modern Machine Learning}.
\newblock \bibinfo{journal}{\emph{J. Mach. Learn. Res.}} \bibinfo{volume}{23},
  \bibinfo{number}{1}, Article \bibinfo{articleno}{226} (\bibinfo{year}{2022}).
\newblock
\showISSN{1532-4435}


\bibitem[Gao et~al\mbox{.}(2020)]%
        {gao2020fuzz}
\bibfield{author}{\bibinfo{person}{Xiang Gao}, \bibinfo{person}{Ripon~K Saha},
  \bibinfo{person}{Mukul~R Prasad}, {and} \bibinfo{person}{Abhik
  Roychoudhury}.} \bibinfo{year}{2020}\natexlab{}.
\newblock \showarticletitle{Fuzz testing based data augmentation to improve
  robustness of deep neural networks}. In \bibinfo{booktitle}{\emph{2020
  IEEE/ACM 42nd International Conference on Software Engineering (ICSE)}}.
  IEEE, \bibinfo{pages}{1147--1158}.
\newblock


\bibitem[Gebru et~al\mbox{.}(2021)]%
        {gebru2021datasheets}
\bibfield{author}{\bibinfo{person}{Timnit Gebru}, \bibinfo{person}{Jamie
  Morgenstern}, \bibinfo{person}{Briana Vecchione},
  \bibinfo{person}{Jennifer~Wortman Vaughan}, \bibinfo{person}{Hanna Wallach},
  \bibinfo{person}{Hal~Daum{\'e} Iii}, {and} \bibinfo{person}{Kate Crawford}.}
  \bibinfo{year}{2021}\natexlab{}.
\newblock \showarticletitle{Datasheets for datasets}.
\newblock \bibinfo{journal}{\emph{Commun. ACM}} \bibinfo{volume}{64},
  \bibinfo{number}{12} (\bibinfo{year}{2021}), \bibinfo{pages}{86--92}.
\newblock


\bibitem[Guo et~al\mbox{.}(2017)]%
        {guo2017calibration}
\bibfield{author}{\bibinfo{person}{Chuan Guo}, \bibinfo{person}{Geoff Pleiss},
  \bibinfo{person}{Yu Sun}, {and} \bibinfo{person}{Kilian~Q Weinberger}.}
  \bibinfo{year}{2017}\natexlab{}.
\newblock \showarticletitle{On calibration of modern neural networks}. In
  \bibinfo{booktitle}{\emph{International Conference on Machine Learning}}.
  PMLR, \bibinfo{pages}{1321--1330}.
\newblock


\bibitem[He et~al\mbox{.}(2021)]%
        {he2021automl}
\bibfield{author}{\bibinfo{person}{Xin He}, \bibinfo{person}{Kaiyong Zhao},
  {and} \bibinfo{person}{Xiaowen Chu}.} \bibinfo{year}{2021}\natexlab{}.
\newblock \showarticletitle{AutoML: A survey of the state-of-the-art}.
\newblock \bibinfo{journal}{\emph{Knowledge-Based Systems}}
  \bibinfo{volume}{212} (\bibinfo{year}{2021}), \bibinfo{pages}{106622}.
\newblock


\bibitem[Hendrycks and Gimpel(2017)]%
        {hendrycks2017}
\bibfield{author}{\bibinfo{person}{Dan Hendrycks} {and} \bibinfo{person}{Kevin
  Gimpel}.} \bibinfo{year}{2017}\natexlab{}.
\newblock \showarticletitle{A baseline for detecting misclassified and
  out-of-distribution examples in neural networks}. In
  \bibinfo{booktitle}{\emph{ICLR}}.
\newblock


\bibitem[Hynes et~al\mbox{.}(2017)]%
        {Hynes2017a}
\bibfield{author}{\bibinfo{person}{Nick Hynes}, \bibinfo{person}{D Sculley},
  {and} \bibinfo{person}{Michael Terry}.} \bibinfo{year}{2017}\natexlab{}.
\newblock \showarticletitle{{The Data Linter: Lightweight, Automated Sanity
  Checking for ML Data Sets}}.
\newblock \bibinfo{journal}{\emph{NeurIPS}} (\bibinfo{year}{2017}).
\newblock


\bibitem[Kwiatkowska(2020)]%
        {kwiatkowska2020safety}
\bibfield{author}{\bibinfo{person}{Marta Kwiatkowska}.}
  \bibinfo{year}{2020}\natexlab{}.
\newblock \showarticletitle{Safety and robustness for deep learning with
  provable guarantees}. In \bibinfo{booktitle}{\emph{Proceedings of the 35th
  IEEE/ACM International Conference on Automated Software Engineering}}.
  \bibinfo{pages}{1--3}.
\newblock


\bibitem[Meyer(1992)]%
        {meyer1992applying}
\bibfield{author}{\bibinfo{person}{Bertrand Meyer}.}
  \bibinfo{year}{1992}\natexlab{}.
\newblock \showarticletitle{Applying ``design by contract''}.
\newblock \bibinfo{journal}{\emph{Computer}} \bibinfo{volume}{25},
  \bibinfo{number}{10} (\bibinfo{year}{1992}), \bibinfo{pages}{40--51}.
\newblock


\bibitem[Mitchell et~al\mbox{.}(2019)]%
        {mitchell2019}
\bibfield{author}{\bibinfo{person}{Margaret Mitchell}, \bibinfo{person}{Simone
  Wu}, \bibinfo{person}{Andrew Zaldivar}, \bibinfo{person}{Parker Barnes},
  \bibinfo{person}{Lucy Vasserman}, \bibinfo{person}{Ben Hutchinson},
  \bibinfo{person}{Elena Spitzer}, \bibinfo{person}{Inioluwa~Deborah Raji},
  {and} \bibinfo{person}{Timnit Gebru}.} \bibinfo{year}{2019}\natexlab{}.
\newblock \showarticletitle{Model cards for model reporting}. In
  \bibinfo{booktitle}{\emph{Proceedings of the conference on fairness,
  accountability, and transparency}}. \bibinfo{pages}{220--229}.
\newblock


\bibitem[Moreno-Torres et~al\mbox{.}(2012)]%
        {moreno2012}
\bibfield{author}{\bibinfo{person}{Jose~G. Moreno-Torres},
  \bibinfo{person}{Troy Raeder}, \bibinfo{person}{Rocío Alaiz-Rodríguez},
  \bibinfo{person}{Nitesh~V. Chawla}, {and} \bibinfo{person}{Francisco
  Herrera}.} \bibinfo{year}{2012}\natexlab{}.
\newblock \showarticletitle{A unifying view on dataset shift in
  classification}.
\newblock \bibinfo{journal}{\emph{Pattern recognition}} \bibinfo{volume}{45},
  \bibinfo{number}{1} (\bibinfo{year}{2012}), \bibinfo{pages}{521--530}.
\newblock


\bibitem[Ren et~al\mbox{.}(2019)]%
        {ren2019likelihood}
\bibfield{author}{\bibinfo{person}{Jie Ren}, \bibinfo{person}{Peter~J Liu},
  \bibinfo{person}{Emily Fertig}, \bibinfo{person}{Jasper Snoek},
  \bibinfo{person}{Ryan Poplin}, \bibinfo{person}{Mark Depristo},
  \bibinfo{person}{Joshua Dillon}, {and} \bibinfo{person}{Balaji
  Lakshminarayanan}.} \bibinfo{year}{2019}\natexlab{}.
\newblock \showarticletitle{Likelihood ratios for out-of-distribution
  detection}.
\newblock \bibinfo{journal}{\emph{NeurIPS}} (\bibinfo{year}{2019}).
\newblock


\bibitem[Schelter et~al\mbox{.}(2019)]%
        {schelter2019}
\bibfield{author}{\bibinfo{person}{Sebastian Schelter}, \bibinfo{person}{Stefan
  Grafberger}, \bibinfo{person}{Philipp Schmidt}, \bibinfo{person}{Tammo
  Rukat}, \bibinfo{person}{Mario Kiessling}, \bibinfo{person}{Andrey Taptunov},
  \bibinfo{person}{Felix Biessmann}, {and} \bibinfo{person}{Dustin Lange}.}
  \bibinfo{year}{2019}\natexlab{}.
\newblock \showarticletitle{Differential data quality verification on
  partitioned data}. In \bibinfo{booktitle}{\emph{2019 IEEE 35th International
  Conference on Data Engineering (ICDE)}}. IEEE, \bibinfo{pages}{1940--1945}.
\newblock


\bibitem[Schelter et~al\mbox{.}(2018)]%
        {Schelter2018}
\bibfield{author}{\bibinfo{person}{Sebastian Schelter}, \bibinfo{person}{Dustin
  Lange}, \bibinfo{person}{Philipp Schmidt}, \bibinfo{person}{Meltem Celikel},
  \bibinfo{person}{Felix Biessmann}, {and} \bibinfo{person}{Andreas
  Grafberger}.} \bibinfo{year}{2018}\natexlab{}.
\newblock \showarticletitle{{Automating large-scale data quality
  verification}}.
\newblock \bibinfo{journal}{\emph{Proceedings of the VLDB Endowment}}
  \bibinfo{volume}{11}, \bibinfo{number}{12} (\bibinfo{year}{2018}),
  \bibinfo{pages}{1781--1794}.
\newblock
\showISSN{21508097}


\bibitem[Shahrokni and Feldt(2013)]%
        {shahrokni2013systematic}
\bibfield{author}{\bibinfo{person}{Ali Shahrokni} {and} \bibinfo{person}{Robert
  Feldt}.} \bibinfo{year}{2013}\natexlab{}.
\newblock \showarticletitle{A systematic review of software robustness}.
\newblock \bibinfo{journal}{\emph{Information and Software Technology}}
  \bibinfo{volume}{55}, \bibinfo{number}{1} (\bibinfo{year}{2013}),
  \bibinfo{pages}{1--17}.
\newblock


\bibitem[Shoeibi et~al\mbox{.}(2021)]%
        {shoeibi2021epileptic}
\bibfield{author}{\bibinfo{person}{Afshin Shoeibi} {et~al\mbox{.}}}
  \bibinfo{year}{2021}\natexlab{}.
\newblock \showarticletitle{Epileptic Seizures Detection Using Deep Learning
  Techniques: A Review}.
\newblock \bibinfo{journal}{\emph{International Journal of Environmental
  Research and Public Health}} \bibinfo{volume}{18}, \bibinfo{number}{11}
  (\bibinfo{year}{2021}).
\newblock
\showISSN{1660-4601}


\bibitem[Stol et~al\mbox{.}(2016)]%
        {stol2016grounded}
\bibfield{author}{\bibinfo{person}{Klaas-Jan Stol}, \bibinfo{person}{Paul
  Ralph}, {and} \bibinfo{person}{Brian Fitzgerald}.}
  \bibinfo{year}{2016}\natexlab{}.
\newblock \showarticletitle{Grounded theory in software engineering research: a
  critical review and guidelines}. In \bibinfo{booktitle}{\emph{Proceedings of
  the 38th International Conference on Software Engineering}}.
  \bibinfo{pages}{120--131}.
\newblock


\bibitem[Zhang(2020)]%
        {zhang2020uncertainty}
\bibfield{author}{\bibinfo{person}{Xiyue Zhang}.}
  \bibinfo{year}{2020}\natexlab{}.
\newblock \showarticletitle{Uncertainty-guided testing and robustness
  enhancement for deep learning systems}. In \bibinfo{booktitle}{\emph{2020
  IEEE/ACM 42nd International Conference on Software Engineering: Companion
  Proceedings (ICSE-Companion)}}. IEEE, \bibinfo{pages}{101--103}.
\newblock


\end{thebibliography}


\end{document}